# Extensive Comparison between INRIM and a Secondary Calibration Laboratory using a Multifunction Electrical Calibrator

F. Galliana and M. Lanzillotti

*Abstract*— An accurate and extensive Inter-laboratory comparison between the laboratory for the calibration of multifunction electrical instruments of the National Institute of Metrology Research (INRIM) and a secondary high level electrical calibration laboratory was performed with satisfactory results. The instrument involved in the comparison was a top class multifunction calibrator, chosen for its wide measurement fields and its excellent definability requiring sensitively small uncertainties to calibrate it. The relevancy of this work is that for the first time at INRIM, a ILC involving a grid of about one hundred and thirty measurement points was carried out. This ILC allowed to exhaustively check the measurement capabilities and exploit the measurement techniques of the secondary laboratory. Attention was also paid to individuate the correlated uncertainty components between the two laboratories measurements. The calibrator resulted adequate to verify the capabilities of high level secondary electrical calibration laboratories, better than fixed Standards or than a 8 1/2-digit multimeter.

## I. INTRODUCTION

Since decades, the measurement capabilities in the field of low frequency electrical quantities (DC and AC Voltage, DC and AC current and DC Resistance) among National Measurements Institutes (NMIs) or high level Laboratories have been verified by means of inter-laboratory comparisons (ILCs) on fixed primary electrical Standards as in [1–4]. These Institutes had the competence to correctly disseminate the electrical units towards their working standards and instruments involved in the traceability transfer to secondary Laboratories that operated with significantly higher measurement uncertainties than NMIs. Now, modern electrical secondary calibration Laboratories are equipped with high accuracy and stability digital instrumentation as multimeters (DMMs) and multifunction calibrators (MFCs) operating in wide measurement fields of the low frequency electrical quantities [5] assuring to these Laboratories sensitively better measurement uncertainties than in the past. The calibration of these instruments can be carried out at different uncertainty levels and with different measurement strategies. Two particular instruments can be calibrated by means of the "artifact calibration", an easy method that requires only few reference Standards [6–9]. Another method that allows the calibration in a higher number of points is described in [10, 11]. The reliability of these electrical Laboratories can be guaranteed by their participation with positive results to technically appropriate ILCs. In this paper, an accurate and extensive ILC involving a top-class 8 1/2-digit MFC between the laboratory for calibration of multifunction electrical instruments of the National Institute of Metrology Research (INRIM-Lab) and a high level secondary electrical calibration laboratory (Cal-Lab), is presented.

It was the first time at INRIM in which a ILC with a secondary Laboratory on so many measurement points and involving several measurement techniques in low frequency electrical quantities, was conducted. This kind of ILC allows to check and exploit completely and adequately the measurement capabilities and techniques of high-level secondary laboratories. The Cal-Lab is accredited with sensitively small uncertainties and the ILC was also carried out to verify its competence and capabilities to maintain and improve its accreditation status.

## II. CHOICE OF SUITABLE INSTRUMENT(S) FOR ILCS

For National Measurement Institutes (NMIs) or ILCs providers, the challenge is to have suitable instruments and expertise to provide appropriate ILCs to exhaustively verify the capabilities of secondary Laboratories. ILCs concerning the calibration of only fixed electrical Standards [1–4, 12, 13] unfortunately don't cover the wide operating fields of modern secondary electrical calibration Laboratories. An instrument covering wide fields and used by INRIM for several years in ILCs with secondary Laboratories is the 8 1/2-digit high precision DMM. An ILC with this instrument ensures a appropriate check of the capabilities of medium-high level secondary laboratories. These laboratories are equipped with high accuracy multifunction instruments as reference Standards to be calibrated at NMIs (traceability chain of fig. 1).

The authors are with the Technical Service for Calibration laboratories, Istituto Nazionale di Ricerca Metrologica, Torino 10135, (f.galliana@inrim.it).

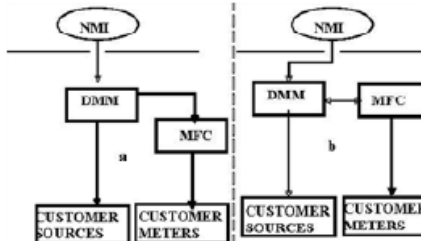

Fig. 1. Traceability transfer from a NMI to medium-high level secondary electrical calibration Laboratories through a high precision DMM. In a) the DMM acts as reference standard, while in b) as transfer standard [5]. Secondary laboratory's customers can be other secondary laboratories operating with worse uncertainties, military Institutions, tertiary or industrial laboratories or their clients.

This figure shows two traceability schemes in which a secondary Laboratory sends to the NMI for calibration only a high accuracy DMM. In the scheme a) the DMM is used as reference Standard for all quantities, and its function is dual because it allows the calibration of the MFC and it will also be used to calibrate customer sources. In the scheme b) the DMM does not act as reference Standard, but as a traceability transfer: it is calibrated by the NMI with particular care and small uncertainties in the measurement points in which the MFC of the Laboratory has to be adjusted. When the DMM comes back to the laboratory it is used to transfer its traceability to the MFC which then will act as Laboratory reference Standard. With the operating mode of the scheme b) the uncertainties with which the MFC can be used are better than in that of the scheme a) and at the same level of those of the DMM. This DMM utilization is well described in [5].

High level secondary calibration Laboratories are instead equipped with complete sets of primary Standards (traceability chain of fig. 2).

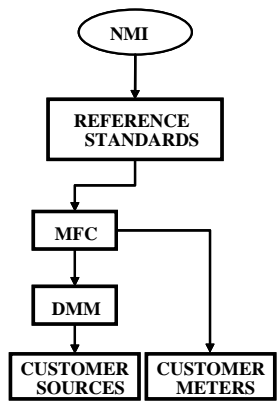

Fig. 2. Traceability transfer from a NMI to a high level secondary electrical calibration laboratory trough a complete set of primary Standards.

Primary Standards are for example a 10 V DC Voltage Standard, DC Voltage dividers, standard resistors and shunts and an AC/DC Voltage transfer Standard. With such instrumentation, these Laboratories can calibrate with considerably small uncertainties their MFC(s) and DMM(s). Till now, the capabilities of these Laboratories have been verified by INRIM by means of ILCs concerning the calibration of a DMM and of some fixed Standards as a 10 V or a 10 kΩ Standards. Nevertheless, these ILCs don't allow to exhaustively verify the capabilities and the measurement uncertainties of these high level Laboratories. In this new ILC, a top-class MFC was chosen for its wide measurement fields and its excellent definability (definitional uncertainty) [14], better than that of a DMM, requiring sensitively small uncertainties to calibrate it. For this reason this kind of ILC can be adequate to check high level electrical Laboratories capabilities.

### III. THE INSTRUMENT TO CALIBRATE IN THE ILC

The instrument under calibration in the ILC was a J. Fluke 5700A MFC with associate a transconductance amplifier J. Fluke 5725A. The operating ranges of this MFC in DC and AC Voltage functions span from 1 mV to 1100 V and at frequencies from 10 Hz to 1.2 MHz in AC Voltage, in Resistance function span from 1 Ω to 100 MΩ, in DC and AC current functions span from 1 µA to 10 A and at frequencies from 10 Hz to 10 kHz for AC current. The accuracy specifications of this MFC are better than those of 81/2 digit DMMs, so with this instrument it is possible to calibrate those DMMs.

*A. ILC instructions*

The calibration of the MFC had to be performed with the instrument in thermal equilibrium with the environment at a temperature of $(23.0 \pm 1)$ °C, after a feeding period of at least 24 h with a sinusoidal voltage of 240 V rms, frequency 50.0 Hz and distortion less than 1%. After the successfully execution of the SELF DIAG and CAL zero procedures, it had to be calibrated in the measurement ranges reported in Table 1. All the measurement points can be seen in the following Tables 2 and 3 and in Fig. 6 to 10.

TABLE I
MEASUREMENT RANGES OF THE ILC WITH THE MFC.

| QUANTITY | MEASUREMENT RANGE | FREQUENCY RANGE |
|---|---|---|
| DC Voltage | 1 mV ÷ 1000 V | |
| AC Voltage | 1 mV ÷ 1000 V | 40 Hz÷1MHz |
| DC Current | 10 µA÷ 10 A | |
| AC Current | 100 µA÷ 10 A | 40 Hz÷5 kHz |
| DC Resistance | 1 Ω ÷ 100 MΩ | |

Normally at the INRIM-Lab a complete calibration of a MFC is performed in three steps [5]. With an initial verification, a wide set of measurement points, in which the MFC operates, are compared with the reference Standards. Successively the adjustment, as suggested by the manufacturer, is performed. A final verification (as performed in the initial one) checks the effectiveness of the adjustment. All the measurement differences between the provided values by the MFC and those measured by the reference system in the two verifications are recorded and inserted in the calibration certificates. For this ILC, only a verification (without adjustment) had to be performed.

## IV. TRACEABILITY TO NATIONAL STANDARDS OF THE INRIM-LAB AND OF THE CAL-LAB

The traceability chain of the INRIM-Lab from National Standards till down to the calibration of the MFC of the ILC is shown in Fig. 3. Among the primary references there are a high precision DMM characterized in linearity and used as DC Voltage ratio standard [15, 16] and a INRIM-made automated DC Voltage fixed ratios divider [17]. The traceability chain of the Cal-Lab is similar to that of the INRIM-Lab with some common Standards also calibrated at INRIM implying a partial correlation between the measurements of the two Laboratories.

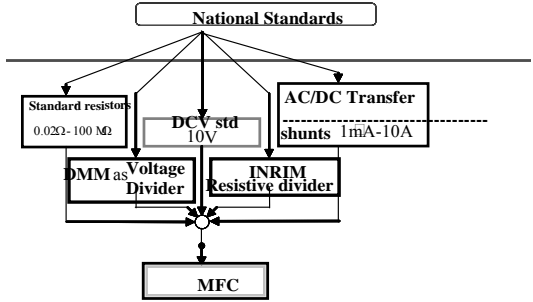

Fig. 3. Traceability chain of the INRIM-lab from National Standards till down to the calibration on the MFC of the ILC.

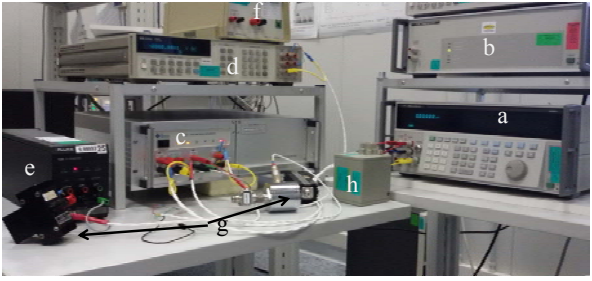

Fig. 4. View of the measurement setup at INRIM-Lab to calibrate the MFC. Are visible: a) the MFC, b) the transconductance amplifier, c) the automated DC Voltage fixed ratios divider, d) the DMM as DC Voltage divider, e) the 10 V Standard, f) a DC Resistance shunt, g) AC/DC Resistance shunts and h) the 10 kΩ Standard. In the photograph are not visible the AC Voltage Standard and the oil and air baths containing respectively the low value and high value standard resistors.

## V. ANALYSIS OF THE RESULTS

The MFC was calibrated twice by the Cal-Lab, once before and once after its calibration at the INRIM-Lab. To minimize the effect of the possible drift of the MFC, the Cal-Lab mean values of its two calibrations were compared with the INRIM-Lab measurements. For the evaluation of the ILC it was considered as measurand the MFC "relative error" defined in the following (1) and (2). For each measurement point, INRIM-Lab and Cal-Lab relative errors were defined respectively as:

$$E_I = (m_I - s)/s \qquad (1)$$

$$E_L = \frac{(m_{L1} - s) + (m_{L2} - s)}{2s} \qquad (2)$$

where $m_I$ indicate the values measured by the INRIM-Lab at the settings $s$ set of the MFC, while $m_{L1}$ and $m_{L2}$ indicate the same values measured by the Cal-Lab before and after its calibration at the INRIM-Lab at the same $s$. For each measurement point, a new measurand as difference between $E_L$ and $E_I$ [12, 13, 18] was introduced.

$$d = E_L - E_I \qquad (3)$$

whose relative standard uncertainty is:

$$u^2(d) = [u^2(E_I) + u^2(E_L) - 2u(E_L)u(E_I) \times r(E_L, E_I)] \qquad (4)$$

where $u(E_L)$ and $u(E_I)$ are respectively the Cal-Lab and INRIM-Lab standard uncertainties, while $r(E_L,E_I)$ is the correlation coefficient between their measurements. Criterions to evaluate $r(E_L,E_I)$ for each electrical quantity were introduced.

. DC Voltage: In each measurement point $r(E_L,E_I)$ was evaluated according to the following relation:

$$r(E_L, E_I) = \frac{u_B^2(std_{\_DCV})}{u(E_I) \times u(E_L)} \qquad (5)$$

where $u_B(std_{\_DCV})$ is the standard uncertainty of the INRIM calibration of the 10 V Standard inserted in the traceability chain of both Laboratories.

$u_B(std_{\_DCV})$ includes the uncertainties respectively due to the DC Voltage National Standard and to the transfer to the INRIM 10 V Standard.

- AC Voltage:
  In each measurement point $r(E_L,E_I)$ was evaluated according to the following relation:

$$r(E_L, E_I) = \frac{u_B^2(std_{\_DCV}) + [u_B(V_{adj}) \times u_B(corr_{load})]}{u(E_I) \times u(E_L)} \qquad (6)$$

where $u_B(corr_{\_load})$ is the type B uncertainty of the correction of the AC Voltage measurements due to the load effects and $u_B(V_{adj})$ is the type B uncertainty of the DC Voltage values involved in the "periodic calibration" at INRIM of the AC/DC transfer Standard [19][a] inserted in the traceability chain of both Laboratories.

- DC Resistance: In each measurement point $r(E_L,E_I)$ was evaluated according to the following relation:

$$r(E_L, E_I) = \frac{u_B^2(std_{RES})}{u(E_I) \times u(E_L)} \qquad (7)$$

where $u_B(std_{RES})$ is the type B standard uncertainty of the calibration at INRIM of the standard resistors of the same values inserted in the traceability chain of both Laboratories.

$u_B(std_{RES})$ includes the uncertainties respectively due to the DC Resistance National Standard and to the transfer to the DC Resistance scale.

- DC current: In each measurement point $r(E_L,E_I)$ was evaluated according to the following relation:

$$r(E_L, E_I) = \frac{u_B(std_{\_DCV}) \times u_B(std_{RES})}{u(E_I) \times u(E_L)} \qquad (8)$$

where $u_B(std_{\_RES})$ is the type B standard uncertainty of the resistors or DC Resistance shunts inserted in the traceability chain of both Laboratories and used to obtain the desired currents.

- AC current. In each measurement point $r(E_L,E_I)$ was evaluated according to the relation:

$$r(E_L, E_I) = \frac{u_B(std_{\_DCV}) \times u_B(corr_{load})]}{u(E_I) \times u(E_L)} \qquad (9)$$

---

[a] The two Laboratories used the same AC/DC transfer Standard but with different release.

as the common AC/DC Resistance shunts are not calibrated, but only verified at INRIM to respect their specifications.

Finally, the normalized error $E_n$ with respect to INRIM-Lab for each measurement point was evaluated as:

$$E_n = \frac{d}{U(d)} \quad (10)$$

where $U(d) = 2u(d)$ at a 95% confidence level.

In Table 2 and 3 the results for DC and AC Voltage are respectively reported. The uncertainties of the two Laboratories are respectively evaluated in calibration procedures respectively approved by INRIM as signatory of the CIPM MRA[b] and by the Italian calibration Accreditation body ACCREDIA. The uncertainty components taken into account by the INRIM-Lab were evaluated according to the criteria defined in [5]. INRIM-Lab and Cal-Lab calibration procedures take into account, for AC measurements, of the output impedance of the calibrator, of the impedances of cables and connectors and of the input impedance of the standard measurement systems. This job was made according to the expertise of the two Laboratories and to the manufacturer indications. For this reason, the AC measurement results are corrected ones due to load effects and, in their uncertainties, a component due to the made corrections is inserted.

---

[b] The CIPM Mutual Recognition Arrangement (CIPM MRA) is the framework through which National Metrology Institutes demonstrate the international equivalence of their measurement standards and the calibration and measurement certificates they issue. The outcomes of the Arrangement are the internationally recognized (peer-reviewed and approved) Calibration and Measurement Capabilities (CMCs) of the participating institutes.

TABLE II
RESULTS OF THE ILC FOR DC VOLTAGE. ALL THE UNCERTAINTIES ARE REPORTED AT 1 σ CONFIDENCE LEVEL.

| Set value (mV) | $E_I$ (×10⁻⁶) | $u(E_I)$ (×10⁻⁶) | $E_L$ (×10⁻⁶) | $u(E_L)$ (×10⁻⁶) | $d$ (×10⁻⁶) | $u_B(std\_DCV)$ (×10⁻⁶) | $u(d)$ (×10⁻⁶) | $E_n$ |
|---|---|---|---|---|---|---|---|---|
| 1 | −25.0 | 96 | 50.0 | 155 | 75.0 | 0.1 | 182.3 | 0.2 |
| −1 | −20.0 | 96 | −50.0 | 155 | −30.0 | 0.1 | 182.3 | −0.1 |
| 3 | 5.0 | 32.5 | 16.7 | 55 | 11.7 | 0.1 | 63.9 | 0.1 |
| 10 | 5.5 | 11.5 | 5.0 | 24 | −0.5 | 0.1 | 26.6 | 0.0 |
| −10 | 4.0 | 11.5 | −5.0 | 24 | −9.0 | 0.1 | 26.6 | −0.2 |
| 100 | 0.5 | 1.1 | 2.5 | 1.8 | 2.0 | 0.1 | 2.1 | 0.5 |
| −100 | 0.5 | 1.1 | 0.0 | 1.8 | −0.5 | 0.1 | 2.1 | −0.1 |
| (V) | | | | | | | | |
| 0.3 | 0.5 | 0.9 | 1.5 | 1.7 | 1.0 | 0.1 | 1.9 | 0.3 |
| −0.3 | 0.2 | 0.9 | −0.3 | 1.7 | −0.5 | 0.1 | 1.9 | −0.1 |
| 1 | −0.2 | 0.4 | 0.7 | 0.5 | 0.9 | 0.1 | 0.6 | 0.7 |
| −1 | 0.0 | 0.4 | 1.1 | 0.5 | 1.1 | 0.1 | 0.6 | 0.8 |
| 3 | 2.3 | 0.5 | 2.7 | 0.6 | 0.4 | 0.1 | 0.8 | 0.3 |
| −3 | 2.3 | 0.5 | 1.8 | 0.6 | −0.5 | 0.1 | 0.8 | −0.3 |
| 7 | 2.3 | 0.4 | 2.2 | 0.4 | −0.1 | 0.1 | 0.6 | −0.1 |
| 10 | 2.2 | 0.3 | 2.1 | 0.5 | −0.1 | 0.1 | 0.5 | −0.1 |
| −10 | 2.3 | 0.3 | 2.1 | 0.5 | −0.2 | 0.1 | 0.5 | −0.2 |
| 20 | 2.5 | 0.4 | 1.5 | 0.6 | −1.0 | 0.1 | 0.7 | −0.7 |
| −20 | 2.5 | 0.4 | 2.0 | 0.6 | −0.5 | 0.1 | 0.7 | −0.3 |
| 30 | 1.7 | 0.5 | 0.9 | 0.8 | −0.8 | 0.1 | 0.9 | −0.4 |
| −30 | 1.7 | 0.5 | 0.9 | 0.8 | −0.8 | 0.1 | 0.9 | −0.4 |
| 50 | 1.5 | 0.6 | 0.6 | 0.6 | −0.9 | 0.1 | 0.8 | −0.6 |
| 100 | 1.2 | 0.4 | 0.8 | 0.6 | −0.4 | 0.1 | 0.7 | −0.3 |
| −100 | 1.2 | 0.4 | 1.1 | 0.6 | −0.1 | 0.1 | 0.7 | −0.1 |
| 300 | 1.7 | 0.7 | 1.3 | 0.7 | −0.4 | 0.1 | 1.0 | −0.2 |
| −300 | 1.6 | 0.7 | 1.0 | 0.7 | −0.6 | 0.1 | 1.0 | −0.3 |
| 400 | 1.8 | 0.7 | 0.9 | 0.7 | −0.9 | 0.1 | 1.0 | −0.5 |
| 800 | 1.5 | 0.5 | 0.7 | 0.6 | −0.8 | 0.1 | 0.8 | −0.5 |
| 1000 | 1.2 | 0.5 | 0.5 | 0.6 | −0.7 | 0.1 | 0.7 | −0.5 |
| −1000 | 1.1 | 0.5 | 1.0 | 0.6 | −0.1 | 0.1 | 0.7 | −0.1 |

TABLE III
RESULTS OF THE ILC FOR AC VOLTAGE. ALL THE UNCERTAINTIES ARE REPORTED AT 1 σ CONFIDENCE LEVEL.

| Set value (mV) | f (kHz) | $E_I$ (×10⁻⁶) | $u(E_I)$ (×10⁻⁶) | $E_L$ (×10⁻⁶) | $u(E_L)$ (×10⁻⁶) | $d$ (×10⁻⁶) | $u_B(std\_DCV)$ (×10⁻⁶) | $u_B(V_{adj})$ (×10⁻⁶) | $u(d)$ (×10⁻⁶) | $E_n$ |
|---|---|---|---|---|---|---|---|---|---|---|
| 1.0 | 1 | 843.7 | 487 | 100.0 | 1300 | −743.7 | 0.1 | 96 | 1381.6 | −0.3 |
| 10.0 | 1 | 59.8 | 52 | −15.0 | 140 | −74.8 | 0.1 | 11.5 | 148.5 | −0.3 |
| 100 | 0.04 | 10.0 | 19 | 0.0 | 30 | −10.0 | 0.1 | 1 | 35.5 | −0.1 |
| 100 | 1 | 0.6 | 19 | −10.0 | 30 | −10.6 | 0.1 | 1 | 35.5 | −0.2 |
| 10 | 10 | −1.5 | 19 | −10.0 | 30 | −8.5 | 0.1 | 1 | 35.5 | −0.1 |
| 200 | 1 | 2.6 | 19 | 0.0 | 27.5 | −2.6 | 0.1 | 1 | 33.1 | 0.0 |
| (V) | | | | | | | | | | |
| 0.3 | 1 | 17.1 | 18 | 1.7 | 23.3 | −15.4 | 0.1 | 0.9 | 29.4 | −0.3 |
| 0.5 | 1 | 5.3 | 15 | −5.0 | 22.0 | −10.3 | 0.1 | 0.9 | 26.3 | −0.2 |
| 1 | 0.04 | 14.0 | 14 | 4.0 | 22.0 | −10.0 | 0.1 | 0.6 | 26.1 | −0.2 |
| 1 | 1 | 5.7 | 14 | −3.0 | 22.0 | −8.7 | 0.1 | 0.6 | 26.1 | −0.2 |
| 1 | 100 | −43.0 | 20 | −35.0 | 55.0 | 8.0 | 0.1 | 0.6 | 58.3 | 0.1 |
| 1 | 300 | −69.2 | 61 | 20.0 | 145.0 | 89.2 | 0.1 | 0.6 | 157.1 | 0.3 |
| 1 | 1000 | −1207.8 | 178 | −550 | 750 | 656.8 | 0.1 | 0.6 | 770.7 | 0.4 |
| 2 | 1 | 5.3 | 14 | 1.8 | 22.5 | −3.5 | 0.1 | 0.6 | 26.5 | −0.1 |
| 3 | 0.04 | 15.0 | 14 | 23.3 | 21.7 | 8.3 | 0.1 | 0.4 | 25.8 | 0.2 |
| 3 | 1 | 12.8 | 14 | 15.0 | 21.7 | 2.2 | 0.1 | 0.4 | 25.8 | 0.0 |
| 3 | 100 | 129.3 | 23 | 158.3 | 21.7 | 29.0 | 0.1 | 0.4 | 31.6 | 0.5 |
| 6 | 1 | −2.1 | 14 | 10.8 | 25.0 | 12.9 | 0.1 | 0.4 | 28.4 | 0.2 |
| 10 | 0.04 | 13.1 | 14 | 20.0 | 20.0 | 6.9 | 0.1 | 0.4 | 24.1 | 0.1 |
| 10 | 1 | 4.5 | 14 | 15.0 | 20.0 | 10.5 | 0.1 | 0.4 | 24.1 | 0.2 |
| 10 | 20 | 2.8 | 14 | 10.0 | 20.0 | 7.2 | 0.1 | 0.4 | 24.1 | 0.1 |
| 10 | 100 | −47.2 | 23 | −10.0 | 50.0 | 37.2 | 0.1 | 0.4 | 55.0 | 0.3 |
| 10 | 300 | −110.6 | 136 | −45.0 | 125.0 | 65.6 | 0.1 | 0.4 | 184.7 | 0.2 |
| 10 | 1000 | −1368.5 | 324 | −850.0 | 750.0 | 518.5 | 0.1 | 0.4 | 817.0 | 0.3 |
| 20 | 1 | 5.0 | 14 | 20.0 | 20.0 | 15.0 | 0.1 | 0.4 | 24.4 | 0.3 |
| 30 | 0.04 | 20.6 | 18 | 30.0 | 21.7 | 9.4 | 0.1 | 0.7 | 28.2 | 0.2 |
| 30 | 1 | 16.6 | 18 | 28.3 | 21.7 | 11.8 | 0.1 | 0.7 | 28.2 | 0.2 |
| 30 | 100 | 120.0 | 50 | 156.7 | 58.3 | 36.6 | 0.1 | 0.7 | 76.8 | 0.2 |
| 60 | 1 | 8.5 | 18 | 200 | 20.8 | 11.5 | 0.1 | 0.7 | 27.5 | 0.2 |
| 100 | 0.04 | 25.5 | 18 | 300 | 21.0 | 4.5 | 0.1 | 0.8 | 27.6 | 0.1 |
| 100 | 1 | 17.1 | 18 | 250 | 21.0 | 7.9 | 0.1 | 0.8 | 27.6 | 0.1 |
| 100 | 20 | 7.0 | 18 | 200 | 21.0 | 13.0 | 0.1 | 0.8 | 27.6 | 0.2 |
| 100 | 100 | −38.0 | 50 | −55 | 60.0 | −17.0 | 0.1 | 0.8 | 78.1 | −0.1 |
| 200 | 1 | 17.0 | 19 | 350 | 20.0 | 18.0 | 0.1 | 0.8 | 27.2 | 0.3 |
| 300 | 0.04 | 6.0 | 20 | 200 | 21.7 | 14.0 | 0.1 | 0.7 | 29.5 | 0.2 |
| 300 | 1 | 11.2 | 20 | 200 | 21.7 | 8.8 | 0.1 | 0.7 | 29.5 | 0.1 |
| 300 | 20 | −3.3 | 20 | 11.67 | 25.0 | 15.0 | 0.1 | 0.7 | 32.0 | 0.2 |
| 600 | 1 | 1.1 | 23 | 12.50 | 25.0 | 11.4 | 0.1 | 0.7 | 33.6 | 0.2 |
| 1000 | 0.04 | 3.1 | 29 | 100 | 25.0 | 6.9 | 0.1 | 0.7 | 37.9 | 0.1 |
| 1000 | 1 | 6.9 | 29 | 200 | 25.0 | 13.1 | 0.1 | 0.7 | 37.9 | 0.2 |
| 1000 | 20 | −19.1 | 45 | 0.0 | 80.0 | 19.1 | 0.1 | 0.7 | 91.8 | 0.1 |
| 1000 | 30 | −26.2 | 80 | −10 | 80.0 | 16.2 | 0.1 | 0.7 | 113.1 | 0.1 |

In Figures from 5 to 9 the *En* values of the ILC in all the measurement points of the ranges of Table 1 are shown.

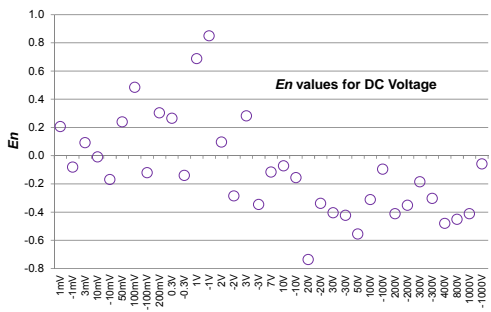

Fig. 5. *En* values for DC Voltage.

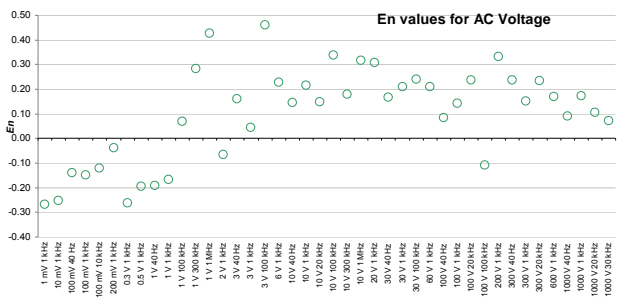

Fig. 6. *En* values for AC Voltage.

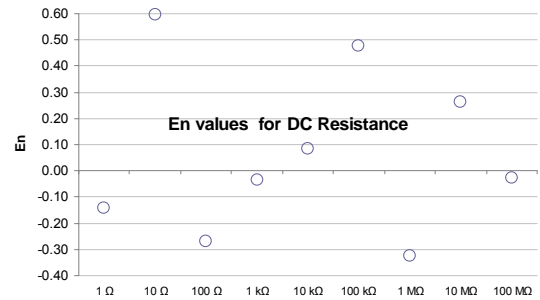

Fig. 7. *En* values for DC Resistance.

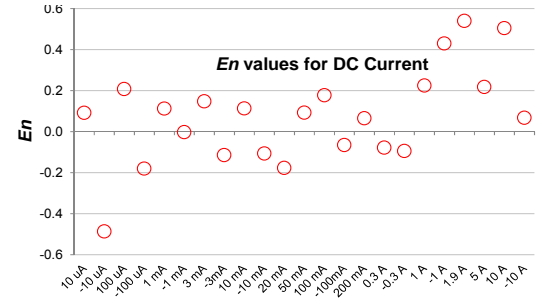

Fig. 8. *En* values for DC Current.

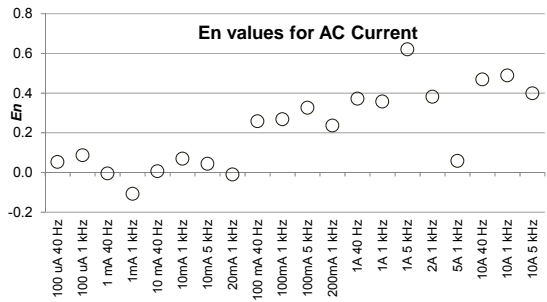

Fig. 9. *En* values for AC Current.

From these graphs it can be seen that $|E_n|$ was less than 1 for each measurement point of the ILC. The $|E_n|$ mean value was 0.3 for DC Voltage and DC Resistance while it was 0.2 for the other quantities.

## CONCLUSIONS

The result of the ILC can be considered satisfactory as the Cal-Lab operate with very small uncertainties although it is a secondary Laboratory. The definite criterions to evaluate the correlation coefficient between the measurements of the two Laboratories can be more useful for ILCs in which the uncertainties of the correlated terms are higher than in this case. The Cal-Lab demonstrated to have adequate competence, instrumentation and calibration procedures to sustain its capabilities. The used 81/2-digit MFC resulted eligible to exhaustively evaluate the capabilities of high level secondary electrical calibration Laboratories for its wide measurement fields and its excellent definability. In addition, it showed high stability and insensibility to transport during the comparison to be considered suitable also for multilateral ILCs that INRIM is actually carrying out with the best accredited secondary laboratories.

## ACKNOWLEDGMENT

The authors wish to thank the ARO S.r.l. that made available the MFC for the comparison and their technicians. The authors greatly thank G. La Paglia, INRIM former technician, to whom is mainly due the development of the INRIM-Lab since the late eighties. The project and evaluation of this ILC was one of his last efforts for INRIM electrical metrology before retirement.

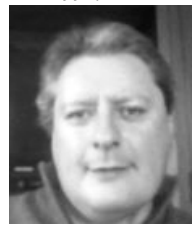

**Flavio Galliana** was born in Pinerolo, Italy, in1966. He received the M.S. degree in physics from the Università degli Studi di Torino, Torino, Italy, in 1991.In 1993 he joined the Istituto Elettrotecnico Nazionale "Galileo Ferraris (IEN), Torino, where he was involved in precision high resistance measurements. He also joined the "Accreditation of Laboratories" Department of IEN. From 2001 to 2005 he was responsible of the Accreditation of Laboratories" Department of IEN. Since 2006, being IEN part of the National Institute of Metrological Research (INRIM), Torino, he has been involved in precision resistance measurements and recently in ILCs technical managements.

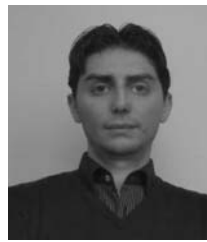

**Marco Lanzillotti** was born in Turin, Italy, in 1981. He received the technical school degree in electronic and telecommunications from I.T.I.S. "E. Majorana", Grugliasco, Torino, in 2001. In 2002 He joined the Electrical Metrology Department of the Istituto Elettrotecnico Nazionale "Galileo Ferraris", now Istituto Nazionale di Ricerca Metrologica (I.N.RI.M.), Turin, in 2002. His main activity has been for researches about the metrological characteristics of the high precision multifunction instruments and the ac-dc transfer in voltage e current and for the calibration of such instruments. Since 2010 he has also been involved in the activity for accreditation and auditing external laboratories.